\documentclass[article,twocolumn,showpacs,preprintnumbers,amsmath,amssymb, superscriptaddress]{revtex4-1}
\usepackage{graphicx}
\usepackage{dcolumn}
\usepackage{bm}
\usepackage{fancyhdr}
\usepackage{float}
\usepackage{ifthen}
\usepackage{eurosym}
\usepackage{wrapfig}
\usepackage{xfrac}
\usepackage{color}

\usepackage[normalem]{ ulem }
\usepackage{soul}

\begin{document}
\title{Spin resonance and magnetic order in an unconventional superconductor}

\ \author{D.~G.~Mazzone}
\ \affiliation{Laboratory for Neutron Scattering and Imaging, Paul Scherrer Institut, 5232 Villigen PSI, Switzerland}

\ \author{S.~Raymond}
\altaffiliation{raymond@ill.fr}
\ \affiliation{Univ. Grenoble Alpes and CEA, INAC, MEM, F-38000 Grenoble, France}

\ \author{J.~L.~Gavilano}
\ \affiliation{Laboratory for Neutron Scattering and Imaging, Paul Scherrer Institut, 5232 Villigen PSI, Switzerland}

\ \author{P. Steffens}
\ \affiliation{Institut Laue-Langevin, 38042 Grenoble, France}

 \author{A. Schneidewind}
\ \affiliation{J\"ulich Center for Neutron Science JCNS, Forschungszentrum J\"ulich GmbH, Outstation at MLZ, D-85747 Garching, Germany}

\ \author{G.~Lapertot}
\ \affiliation{Univ. Grenoble Alpes and CEA, INAC, PHELIQS, F-38000 Grenoble, France}

\ \author{M.~Kenzelmann}
\ \affiliation{Laboratory for Scientific Developments and Novel Materials, Paul Scherrer Institut, 5232 Villigen PSI, Switzerland}

\date{\today}
             
\begin{abstract}
Unconventional superconductivity in many materials is believed to be mediated by magnetic fluctuations. It is an open question how magnetic order can emerge from a superconducting condensate and how it competes with the magnetic spin resonance in unconventional superconductors. Here we study a model $d$-wave superconductor that develops spin-density wave order, and find that the spin resonance is unaffected by the onset of static magnetic order. This result suggests a scenario, in which the resonance in Nd$_{0.05}$Ce$_{0.95}$CoIn$_5$ is a longitudinal mode with fluctuating moments along the ordered magnetic moments.

\end{abstract}
\maketitle

Unconventional superconductors feature superconducting gap functions whose symmetry is lower than the $s$-wave gap of conventional superconductors \cite{BCS1957}. In addition to the Gauge symmetry, an unconventional superconductor breaks the point and/or the time reversal symmetry of the normal state. As a consequence, the orbital part of the superconducting wave-function is described by higher-order spherical harmonics featuring nodal lines or points in reciprocal space where low-energy quasi-particles are present. 

Although the microscopic mechanism for unconventional superconductivity is still unclear, there is increasing evidence that it arises from Cooper pairs mediated by magnetic fluctuations \cite{Monthoux2007}. The intertwined nature of electronic and magnetic degrees of freedom can lead to novel quantum phases, in which magnetic order and superconductivity coexist \cite{Michel2017, Pfleiderer2009, White2015, Norman2011}. The nodal areas probably play an important role in such phenomena. The investigation of such ground states can be an important tool to understand the fundamental interactions that lead to non-phonon-driven superconductivity.

In a number of unconventional superconductors a collective spin-1 excitation is observed below the critical temperature, $T_c$, that appears to be related to magnetically-driven superconductivity. This so-called spin resonance is visible in the dynamical magnetic response probed by inelastic neutron scattering below $T_c$ and has been observed in a variety of unconventional $d$- and $f$-electron systems, such as cuprates, heavy-fermion and iron-based superconductors \cite{Miyake1986, Moriya2000, Stock2008}. A widespread interpretation of the spin resonance derives from the exciton model, where it arises as a feedback effect of superconductivity when the electronic structure acquires a gap. The resonance corresponds to a spin exciton of the superconducting Cooper pairs similar to an electron-hole exciton in an insulator \cite{Bulut1992, Eremin2008}. Alternatively, the resonance has also been associated with a magnon-like excitation of the hybridized $f$-electrons, which is heavily damped above $T_c$. The condensation of the hybridized bands, however, opens a superconducting gap that suppresses the Landau overdamping of the magnetic excitation \cite{Morr1998, Chubokov2008, Song2016}. Regardless of this ongoing debate, the energy of the resonance scales in a universal linear relation with the orbital part of the superconducting gap, providing evidence for a deep connection between antiferromagnetic fluctuations and unconventional superconductivity \cite{Gu2009}. 

In this letter we aim to experimentally answer the open question of what happens to antiferromagnetic fluctuations when static magnetic order develops inside the superconducting condensate. The opening of a spin-density wave (SDW) gap should break the rotational spin symmetry and stabilize a magnetic ground state with a modified excitation spectrum. The change of the spin resonance, as magnetic order arises, may help to understand the microscopic origin of unconventional superconductivity.

The series Nd$_{1-x}$Ce$_{x}$CoIn$_5$ features a coexistence between itinerant magnetism and superconductivity for $x$ $>$ 80\% and is thus a good model system to study the interplay between magnetic order and superconductivity \cite{Hu2008}. The undoped compound, CeCoIn$_5$, is an unconventional superconductor below $T_c$ = 2.3 K with a well-established $d_{x^2-y^2}$-gap symmetry \cite{Petrovic2001, Allan2013, Zhou2013}. Within this superconducting condensate, a spin resonance emerges near the wave-vector $\vec{Q}_{SR}$ = (0.5, 0.5, 0.5) and has an energy of $\Delta E$ = 0.6 meV (3$k_BT_c$) \cite{Stock2008}. Polarized neutron scattering results suggest that the resonance features Ising like moment fluctuations oriented perpendicular to the tetragonal plane \cite{Raymond2015}. 

Substitution of localized Nd$^{+3}$ atoms for Ce in CeCoIn$_5$ strengthens the local moment magnetism, and triggers magnetic order within a heavy-fermion ground state \cite{Hu2008}. At the same time, a small Nd concentration in Nd$_{1-x}$Ce$_x$CoIn$_5$ also acts as source of random disorder that weakens the superconducting pairing strength while most probably preserving the $d_{x^2-y^2}$-gap symmetry \cite{Hu2008, Raymond20142, Mazzone2017}. In Nd$_{0.05}$Ce$_{0.95}$CoIn$_5$ static magnetic order is found below $T_N$ = 0.8 K  within its superconducting phase below $T_c$ = 1.8 K \cite{Hu2008, Raymond20142}. This magnetic structure is an amplitude modulated SDW described by $\vec{Q}_{IC}$ = ($q$, $q$, 0.5) with $q$ = 0.448 and ordered magnetic moment $\mu\approx$ 0.11$\mu_B$ along the $c$-axis.

Here we study the low-energy excitation spectrum of Nd$_{0.05}$Ce$_{0.95}$CoIn$_5$ and observe a spin resonance. The spin resonance is only observed for temperatures below $T_c$, and has an energy $\Delta E$ = 0.43(2) meV. It is thus similar to the spin resonance in undoped CeCoIn$_5$ and shares many of its properties with the spin resonance of other unconventional superconductors. We study its temperature dependence while magnetic order emerges.

Preliminary inelastic neutron scattering data were collected on the cold triple-axis spectrometer PANDA at the Heinz Maier-Leibnitz Zentrum in Munich, Germany. The sample consisted of about 50 co-aligned single crystals. It was observed that the variation of the Nd concentration that is obtained in a multi-assembly of single crystals results in a distribution of magnetic order with identical $T_N$. Thus, all the data shown in this work were carried out in a subsequent experiment with one well-characterized crystal. We used the same single crystal as in Ref. \cite{Raymond20142, Mazzone2017} with $m$ = 64 mg that was oriented within the scattering plane spanned by (1, 1, 0) und (0, 0, 1) in reciprocal lattice units (r.l.u.). The high flux of the cold triple axis spectrometer ThALES at the Institut Laue-Langevin in Grenoble, France provides a unique opportunity to carry out such an inelastic neutron scattering experiment. The initial neutron beam was diffracted from a silicium monochromator, Si (111), and higher order contamination were suppressed by a velocity selector. The spectrometer was used in a $W$ configuration and with a horizontal and vertical focussing option. The outgoing beam was diffracted on a double focussing pyrolitic graphite analyzer with a fixed $k_f$ = 1.45 \AA$^{-1}$ and collimated in front of the detector by means of a radial collimator. 

\begin{figure}[tbh]
\includegraphics[width=\linewidth]{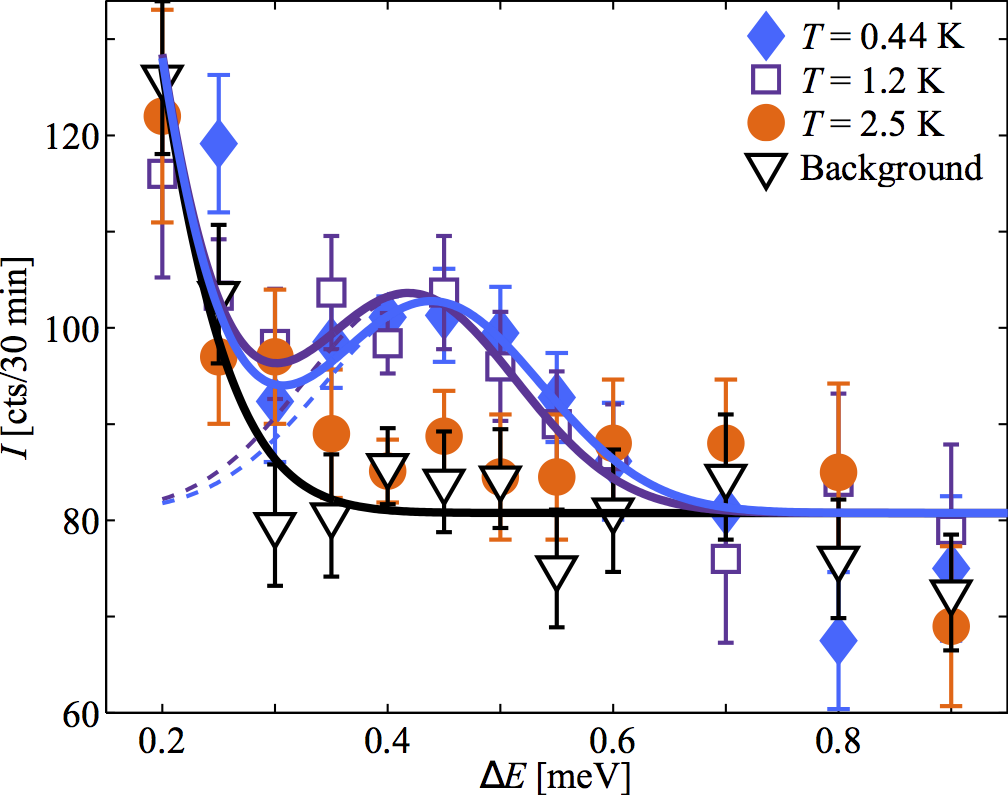}
\caption{(Color online) Neutron excitation spectrum measured at the reciprocal wave-vector $\vec{Q}_{SR}$ = (0.5, 0.5, 0.5) and $T$ = 0.44, 1.2 and 2.5 K. The background denotes the average intensity from the reciprocal lattice positions (0.41, 0.41, 0.83) and (0.54, 0.54, 0.08) measured at $T$ = 0.44 K.}
\label{fig1}
\end{figure}
\bigskip

The low-energy neutron excitation spectrum of Nd$_{0.05}$Ce$_{0.95}$CoIn$_5$ is displayed in Fig. \ref{fig1}. The spectrum was measured at $\vec{Q}_{SR}$ for temperatures $T$ = 0.44, 1.2 and 2.5 K. The background was estimated at $T$ = 0.44 K and at reciprocal wave-vectors sufficiently far from $\vec{Q}_{SR}$, as the scattering in the normal phase may feature paramagnetic scattering. The background shown here was obtained from the average intensity measured at the reciprocal lattice positions (0.41, 0.41, 0.83) and (0.54, 0.54, 0.08). The spectrum reveals a well-defined excitation at energy transfer $\Delta E$ = 0.43(2) meV, which is suppressed at $T$ = 2.5 K $>$ $T_c$. 

\begin{figure}[tbh]
\includegraphics[width=\linewidth]{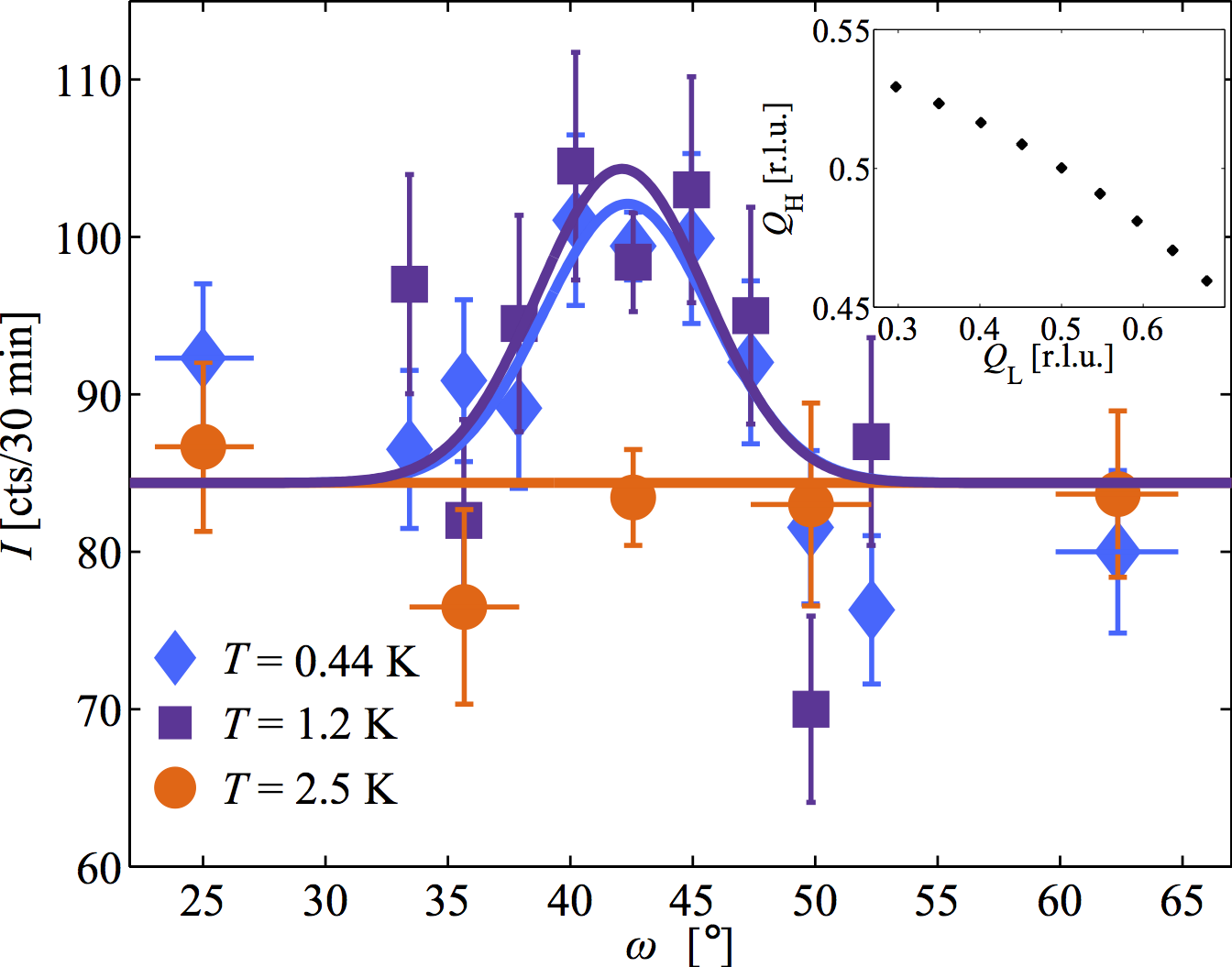}
\caption{(Color online) Rocking $\omega$-scans measured at $\Delta E$ = 0.4 meV for $T$ = 0.44, 1.2 and 2.5 K. The inset displays the corresponding cut in ($Q_H$, $Q_H$, $Q_L$).}
\label{fig2}
\end{figure}

Fig. \ref{fig2} shows rocking $\omega$-scans at constant energy transfer $\Delta E$ = 0.4 meV for temperatures $T$ = 0.44, 1.2 and 2.5 K. Here, $\omega$ denotes the relative rotation angle around the vertical axis that is perpendicular to the scattering plane. A  scan centered around $\vec{Q}_{SR}$  traces the path in reciprocal space shown in the inset of Fig. \ref{fig2}. The data reveal a well-defined excitation at $\omega$ = 42(1)$^\circ$ that vanishes above $T_c$. Within the resolution of our experiment it was not possible to determine whether the excitation is attributed to $\vec{Q}_{SR}$ or to the incommensurate wave-vector $\vec{Q}_{IC}$.

\begin{figure}[tbh]
\includegraphics[width=\linewidth]{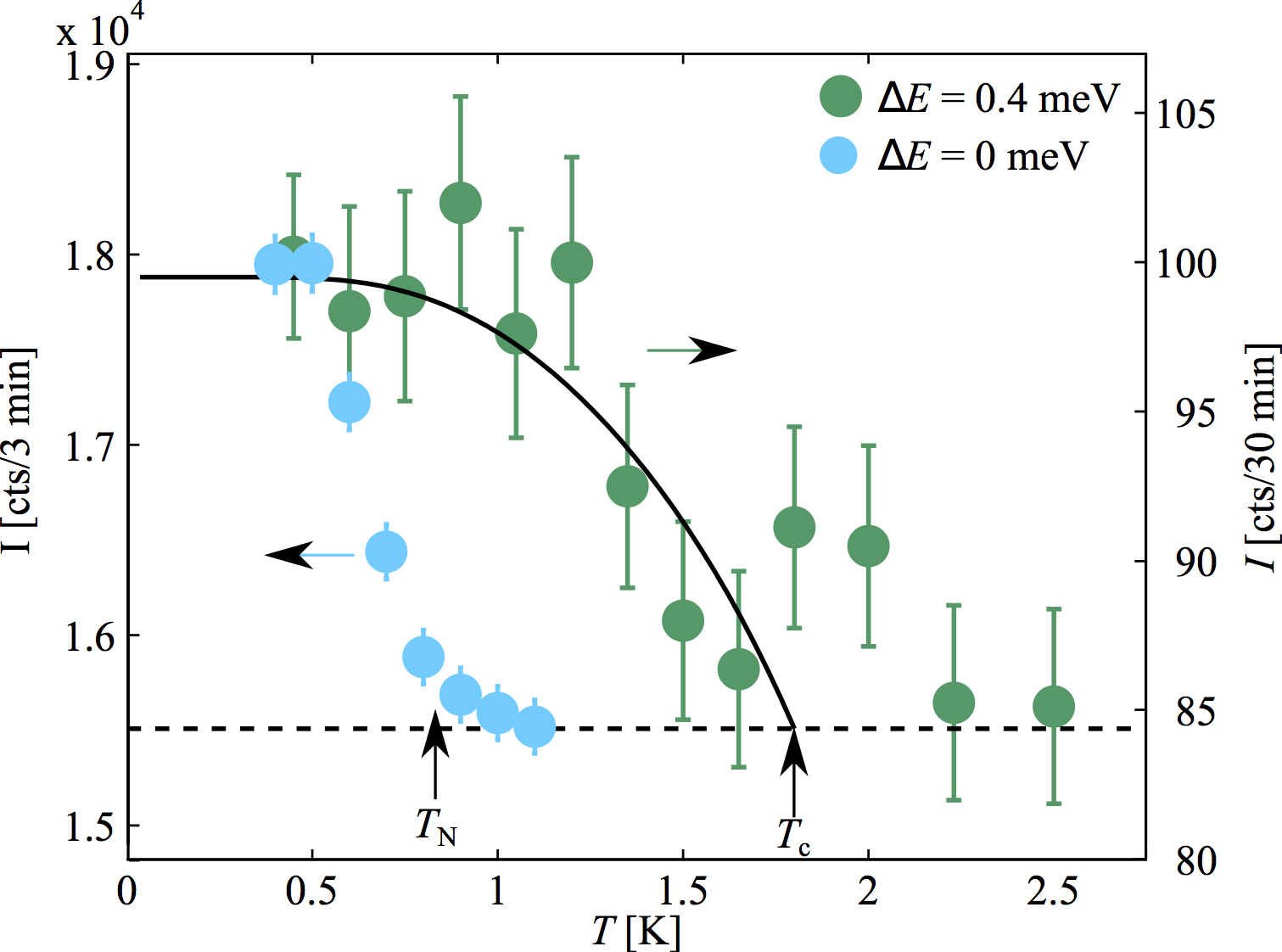}
\caption{(Color online) Temperature dependence of the excitation measured at $\Delta E$ = 0.4 meV and $\vec{Q}_{SR}$ = (0.5, 0.5, 0.5) over plotted with $\Delta E$ = 0 meV and $\vec{Q}_{IC}$ = (0.55, 0.55, 0.5) (taken from Ref. \cite{Raymond20142}). The solid line represents a BCS-fit as guide line to the eye. The dashed line shows the background that was obtained self-consistently from this temperature scan and the $\omega$-scan at $T$ = 2.5 K shown in Fig. \ref{fig2}. Black arrows indicate the superconducting and antiferromagnetic transition temperature.}
\label{fig3}
\end{figure}

The temperature dependence of the scattering at $\Delta E$/$k_BT_c$ = 0.4 meV and $\vec{Q}_{SR}$ is displayed in Fig. \ref{fig3}, compared to the intensity of the magnetic Bragg peak at $\vec{Q}_{IC}$ (taken from Ref. \cite{Raymond20142}). With increasing temperature the excitation decreases in strength and is continuously suppressed around $T_c$ = 1.8 K. The dashed line represents the background that was obtained self-consistently from this temperature scan and the $\omega$-scan at $T$ = 2.5 K shown in Fig. \ref{fig2}. The solid line in Fig. \ref{fig3} describes the temperature dependence of the gap in the BCS (Bardeen-Cooper-Schrieffer) theory, which features a similar temperature dependence. The emergence of the excitation at $T_c$ provides strong evidence that it is a spin resonance related to superconductivity. Within the resolution of our experiment, it appears at the same wave-vector as the one of CeCoIn$_5$ \cite{Stock2008, Raymond2015}. 

\bigskip

The ratio of the gap at the lowest temperatures to the transition temperature is $\Delta E$/$k_BT_c$ = 2.8(1), and thus similar to that in CeCoIn$_5$ as well as that in La and Yb-substituted CeCoIn$_{5}$ \cite{Song2016,Panarin2011}. This demonstrates that the excitation is related to superconductivity but not to the SDW. Moreover, the low energy excitation spectrum of 5\% Nd doped CeCoIn$_5$ reveals a spectral weight that is consistent with the one of the pure compound and where a fluctuating moment of $<\mu_{eff}^2>$ $\approx$ 0.38$\mu_B^2$ is found \cite{Raymond2015, Stock2008}. Here, the intensity of the resonance was compared with the weak structural Bragg peak (1,1,0) and cross-checked with similar data on CeCoIn$_5$. The spin resonance in Nd substituted CeCoIn$_5$ is not affected by the static incommensurate antiferromagnetic order at $T_N$ = 0.8 K (see Fig. \ref{fig3}).  In addition it shows that the temperature dependence of the resonance energy and its intensity reveal no change at the transition below $T_N$. The magnetically ordered phase features a long-range ordered moment $\mu$ = 0.11(5)$\mu_B$ with a lower bound of the magnetic correlation lengths, $\xi_a$ $>$ 150 \AA~and $\xi_c$ $>$ 230 \AA, that is at least one order of magnitude larger than the mean distance, $d\approx$ 17 \AA, between the Nd impurities \cite{Mazzone2017, Raymond20142}. This excludes a scenario in which static magnetic order and superconductivity are phase separated. 

\bigskip
When magnetic order arises in a system of magnetic degrees of freedom interacting through isotropic Heisenberg exchange, it breaks the rotational symmetry and leads to a modification of the magnetic excitation spectrum. The hallmark excitation of the breaking of the continuous spin symmetry is the Goldstone mode that emerges with polarization transverse to the magnetic moment direction. Nd$_{0.05}$Ce$_{0.95}$CoIn$_5$ features a SDW with an amplitude modulated structure and moment orientation perpendicular to the tetragonal basal plane \cite{Mazzone2017}. Since the spin resonance remains unchanged at the antiferromagnetic phase boundary it suggests that the observed excitation has no sizable transverse components and is polarized mainly along the direction of the SDW order. This interpretation is supported by recent inelastic neutron scattering studies of CeCoIn$_5$, where the spin resonance was argued to be of Ising nature with fluctuations along the tetragonal axis \cite{Raymond2015}. 

Theoretically the magnetic excitations of unconventional superconductors with antiferromagnetic order has been investigated in the context of  heavy-fermion compounds, iron based superconductors and cuprates \cite{Chang2007, Knolle2011, Weicheng2014, Rowe2012}. Particularly interesting is the case $T_N$ $<$ $T_c$ that is treated in Ref. \cite{Weicheng2014}. Here it was shown that while magnetic order does not affect the longitudinal component of the superconducting spin resonance, the transverse components soften into spin-waves. This model describes three electronic bands using a random phase approximation and is constructed for iron-based superconductors, so it may not be directly applicable to our case for which no similar model has been developed. However we can expect that the general phenomenology at the antiferromagnetic wave-vector, where a sign change of the superconducting gap occurs, remains valid for Nd$_{0.05}$Ce$_{0.95}$CoIn$_5$ although the Fermi topology is different. This theory is consistent with our suggestion that the antiferromagnetic fluctuations associated with the resonance are polarized along the direction of the ordered magnetic SDW moments.

The reverse case, where unconventional superconductivity emerges within an ordered magnetic phase ($T_N$ $>$ $T_c$), has been experimentally investigated, for instance, in the iron pnictide  Ba(Fe$_{0.953}$Co$_{0.047}$)$_2$As$_2$ \cite{Pratt2009, Dai2015}. Here an emergent spin resonance is observed below $T_c$ that reduces the ordered magnetic moment along the orthorhombic $a$-axis. Details of the spin dynamics were measured by polarized inelastic neutron scattering for a similar underdoped concentration \cite{Wasser2016}. In the antiferromagnetic phases three distinct modes are observed and are reminiscent of the spin dynamics of the antiferromagnetic parent compound BaFe$_{2}$As$_{2}$. Below $T_{c}$ the resonances manifest an additional low energy signal in the two transverse channels while the longitudinal mode is also here unaffected by superconductivity. The antithetic behavior of the spectral weight transfer from the magnetic Bragg peak to the transverse component of the spin resonance was interpreted as a signature of the competition between superconductivity and antiferromagnetic order in Co-underdoped BaFe$_2$As$_2$ \cite{Wasser2016, Pratt2009, Dai2015}. 

This is in complete contrast to Nd-doped CeCoIn$_5$ where no effect of static magnetic order on the spin resonance is observed below $T_N$ $<$ $T_c$, and no novel transverse component (Goldstone mode) of the spin resonance in CeCoIn$_5$ has been found yet. The latter may simply be due to the difficulty to observe these excitations, or it may arise from a difference in the relationship between static magnetic order and the spin resonance in these two classes of materials. 

All these experiments underline the robustness of longitudinal magnetic fluctuations in the interplay of magnetic fluctuations associated with superconductivity and SDW order; either for antiferromagnetic fluctuations within the superconducting condensate (as in BaFe$_2$As$_2$) or for the longitudinal mode of the superconducting spin resonance inside the magnetically ordered phase (as suggested in the case of CeCoIn$_5$). It is tempting to point out the analogy between the gapped magnetic excitation spectrum of an unconventional superconductor and a gapped spin liquid, such as the Haldane chain, where novel quantum excitations survive even when three dimensional antiferromagnetic order develops in the presence of quantum coherence. The difference in the Haldane chain is, however, that excitations are polarized in all three spin directions, two of which develop into Goldstone modes when magnetic order develops \cite{Enderle1999}.

At high fields close to $H_{c_2}$, CeCoIn$_5$ and Nd-doped CeCoIn$_5$ feature another SDW-order \cite{Kenzelmann2008, Mazzone2017}. At such fields, the spin resonance is expected to condense into the ground state \cite{Stock2012, Raymond2012, Akbari2012, Michal2011}, leading to a novel phase with a unique interplay between superconductivity and magnetic order. In fact, our experiments suggest that while at low fields the static and dynamic spin fluctuations remain completely uncoupled, the condensation of the longitudinal correlations towards $H_{c_2}$ may lead to a fundamentally novel phase \cite{Mazzone2017}.

In summary, we report the discovery of a spin resonance in Nd$_{0.05}$Ce$_{0.95}$CoIn$_5$ below $T_c$ = 1.8 K near $\vec{Q}_{SR}$ = (0.5, 0.5, 0.5) and at $\Delta E$ = 0.43(2) meV. The resonance gap is slightly smaller than the one observed in CeCoIn$_5$ but has a similar $\Delta E/T_c$ ratio. Within the resolution of our experiment, it appears at the same wave-vector. The key result of our study is that the spin resonance is not affected by the onset of static magnetic order. The comparison with CeCoIn$_5$ and theoretical models for the iron pnictides suggest a longitudinal resonant mode in Nd$_{0.05}$Ce$_{0.95}$CoIn$_5$ with fluctuating moments in the direction of the ordered magnetic moments of the SDW.

We acknowledge Y. Sidis for illuminating discussions. We thank the institut Laue-Langevin and the Heinz Maier-Leibnitz Zentrum for the allocated beam time on ThALES and on PANDA. We thank the Swiss National Foundation (grant No. 200021\_147071 and 200021\_138018).

\def\bibsection{\section*{\refname}}

\end{document}